\documentclass[12pt,a4paper]{article} 

\usepackage{graphicx}
\usepackage{xcolor}

\begin{document}

\title{Cosmic acceleration driven by dark matter self-interactions: a phenomenological treatment}

\author{ A. Kaz\i m \c Caml\i bel
\\ \small 
\\ \small Electrical and Electronics Engineering Department, 
\\ \small Turkish-German University, 34820 Beykoz, Istanbul, Turkey
\\ \small camlibel@tau.edu.tr}

\maketitle

\abstract{We explore the idea that cosmic acceleration may be a byproduct of late-time effects like structure formation in two steps. First, we consider the equation of state for an inhomogeneous cosmic fluid, which may lead to a Gedanken-model for cosmic evolution, where dark matter is strongly self-interacting and stays in a plasma state until late stages of the cosmic evolution. After decoupling, it condensates to super-structures with cosmic voids similar to the current picture of the universe, introducing a negative pressure term in relation to self-interaction strength. Secondly, we carry out a cosmological analysis inspired by this scenario via a phenomenological ansatz that exhibits a transient behavior. In this analysis, we use the recent Type Ia supernova compilation and high redshift quasar data and compare the results to that of $\Lambda$CDM. It turns out that proposed model can solve the quasar Hubble diagram tension.

Keywords: Cosmic acceleration, Large-scale structure, Self-interacting dark matter}

\section{Introduction}

Despite being strongly favored by cosmological data, $\Lambda$CDM --the standard model of cosmology-- still lacks convincing explanations to its two well-known setbacks: (i) {\it the fine-tuning problem}; the low but nonzero value of the observed vacuum energy density in comparison to the prediction coming from quantum field theory \cite{sahni2002cosmological} and (ii) {\it the coincidence problem}; the surprisingly close present values of energy densities for matter and dark energy components in the cosmic fluid, a problem which implies that we live in a very special era in the cosmic lifetime \cite{zlatev1999quintessence}. One can argue whether those problems are relevant from a cosmological point of view or not \cite{velten2014aspects}; however, it is still reasonable to invert this set of problems in an attempt to make sense of the cosmic puzzle of acceleration: It would be {\it pleasing} to come up with a cosmic scenario, where there is no dark energy and the acceleration of the universe is {\it triggered} by an event that took place in the recent cosmic history. 

Cosmic-scale events that we can attribute to late time evolution are scarce, and they are mostly related to structure formation. The first stars are born around z$\sim$15 \cite{wiklind2012first}, causing a reionization period, an effect that we can single out from cosmological observations. A period of nonlinearization and cosmic structure growth, which can be regarded as a still ongoing process, follows reionization. A hierarchy of structures is pretty much observable to our instruments, starting from galaxies that form into clusters and further super-structures and voids of various sizes between them. Distribution of dark matter (DM) is not far from the visible one, according to the weak lensing observations that give large scale distribution of this mysterious component of cosmic fluid \cite{massey2007dark}.

Deviating from cosmological principle and taking this inhomogeneity into account to see if it can be an explanation to observed cosmic acceleration is not a new idea among cosmologists. A fair amount of work argues that backreaction of matter inhomogeneity may be the reason for the observed acceleration \cite{buchert2012backreaction}. Einstein's field equations can be solved in a perturbed background as well, and the deceleration parameter that is also weakly dependent on space in addition to its usual time dependence can be served as an alternative \cite{kolb2006cosmic}.

Nevertheless, the fact that the universe is not {\it exactly} homogeneous or isotropic does not disclaim the idea that the universe is still at least {\it statistically} homogeneous and isotropic at large scales; the probability of deviating from average density is the same for the whole space. It is fair to assume that the cosmic structure/fluid follows a similar void-filament pattern everywhere in the universe. At this point, it is also fair to ask the following questions: Is it possible to propose a cosmic fluid whose inhomogeneous nature is implicitly given in its equation of state and/or come up with an energy density function that at least phenomenologically takes late-time characteristics of the universe into account? Implementing such fluid in a Friedman-Robertson-Walker (FRW) setting and solving Einstein's equations may give hints on the effects of these late-time features in terms of cosmic expansion.

In this paper, we take an approach in this direction. First, we take an equation of state for a fluid that is inhomogeneous but still may be characterized by its macroscopic properties and derive Einstein's equations accordingly. In doing so, we get a glimpse of the effects of additional features like void-size or structural bond-strengths on the cosmic expansion. Then, we use a phenomenological ansatz for the energy density of these effects and compare such contribution to the latest set of cosmological supernova data and high redshift quasar data.

\section{Equation of State for an Inhomogeneous Fluid}

In fluid dynamics literature, an equation of state for a fluid consisting of {\it walls} and {\it voids} was proposed; it is basically the equation of state for dry foam (bubbles with ideal gas in them) \cite{aref2000equation}: 

\begin{equation}
pV+\frac{2}{3}\sigma A=NkT
\label{eqnfoam}
\end{equation}
Here $p$ and $V$ are the total pressure and volume of the system, $\sigma$ is the surface tension on the bubble surfaces, $A$ is the total area of the interfaces between bubbles, $N$ is the number of ideal gas particles, $T$ is the temperature of the gas within the bubbles, and $k$ is the Boltzmann constant. If we adapt this model to cosmology, we may assume that almost all matter is concentrated in thin walls of structure ($N=0$) or alternatively set $T\sim0$. So, it is possible to come up with a negative pressure term in the form of tension in structure walls,

\begin{equation}
p=-\frac{2}{3}\frac{\sigma A}{V}.
\label{foampress}
\end{equation}
Equation \ref{eqnfoam} involves two competing components to the overall pressure: outward, pressure of the gas within the bubbles and inward, tension within the walls. Setting the first one to zero, we are left with an inward, hence negative pressure.
We can think of this negative pressure as the {\it repelling} part of gravity since pressure itself counterintuitively contributes to attraction in general relativity. 
The term ``$\sigma A/V$" can be treated as the {\it surface energy per volume} and will be denoted by $\rho_s$ from now on.

If we solve Einstein's equations with (\ref{foampress}) for the spatially flat case of the FRW metric, the deceleration parameter takes the form,

\begin{equation}
q=\frac{1}{2}\left(1-2\frac{\rho_s}{\rho_c}\right)
\label{dec}
\end{equation}
where $\rho_c$ is the critical density.
One can easily see that if $\rho_s=\rho_c$, i.e., all energy density in the universe is in the form of surface energy, we recover $q=-\frac{1}{2}$, the value for a universe dominated by cosmic branes. 

If we move on without introducing any exotic components like dark energy or cosmic branes, we may want to see if it is convenient to interpret this tension energy as Newtonian gravitational potential energy within the DM structure. For simplicity, let us assume that the matter content is confined in spherical shells. For such configuration gravitational potential energy, $U$ is given by (as in the standart exercise for electrostatics)

\begin{equation}
U=\frac{Gm^2}{2\mathcal{R}}
\end{equation}

where $m$ is the total mass of the shell, $\mathcal{R}$ is the radius and $G$ is the gravitational constant. Rewriting $m$ in terms of surface mass density $\mu$,

\begin{equation}
m=\mu 4\pi \mathcal{R}^2
\end{equation}

we get an estimation of the energy density independent of the void radius,

\begin{equation}
\rho_s=\frac{U}{\mathcal{V}}=G\mu^2 6\pi
\label{rhos}
\end{equation}

We assume that the voids are almost empty, so 

\begin{equation}
\mu c^2\mathcal{A}=\rho_c\mathcal{V}
\label{rhoc}
\end{equation}

where $\mathcal{A}$ and $\mathcal{V}$ are the surface area and volume of the spherical shell, respectively. Plugging (\ref{rhos}) and (\ref{rhoc}) into (\ref{dec}) and rearranging terms, we get the following equation for the deceleration parameter:

\begin{equation}
q=\frac{1}{2}\left(1-\frac{H^2\mathcal{R}^2}{4c^2}\right)
\label{deceleration}
\end{equation}

We can make an estimation for the term with the negative sign to see if it can sustain the current acceleration. The Hubble parameter  can be estimated as $H\sim70$ km/s/Mpc \cite{bennett20141}, and the average void radius from surveys is $\mathcal{R}\sim 100$ Mpc \cite{mao2017cosmic}. It turns out that the introduced contribution is only about $10^{-4}$. We can also calculate the necessary void size for acceleration (e.g., $q=-1/2$), which is about $\mathcal{R}\sim 10^5$ Mpc, bigger than the Hubble horizon itself.

It would be optimistic to expect that a gravity-only tension within the cosmic structures to drive the cosmic acceleration. But we are well aware that the gravity is not the only long-range interaction in the universe, and it is actually the weakest by far. To assume that DM {\it particles} are not interacting with each other is still part of the benchmark cosmology, but this assumption is being heavily argued lately \cite{tulin2018dark}. Actually, it is natural to think that DM particles should be interacting with each other in a yet unknown non-{\it standard model} mechanism, like every other particle in the universe does through some interaction other than gravity.

Once taking self-interacting DM models into account, we would like to rewind the cosmic evolution to identify a past DM plasma stage where the universe was small and too hot for DM to sustain any structure. Such a cosmic dark plasma scenario was considered in the literature \cite{cyr2013cosmology}, but there is no reason to expect such an era to take place before the photon-baryon decoupling. On the contrary, considering that DM is five times denser than the baryonic matter in the universe, it is possible that DM-plasma would decouple much later than photon-baryon plasma, depending on the type and strength of the DM self-interaction itself. Recent observations of early galaxies with no DM can be regarded as hints of yet uncoupled DM-plasma at that time \cite{lang2017falling}. 

\section{Analysis with a Phenomenological Ansatz}

\begin{figure*}
\centering
\includegraphics[width=14cm]{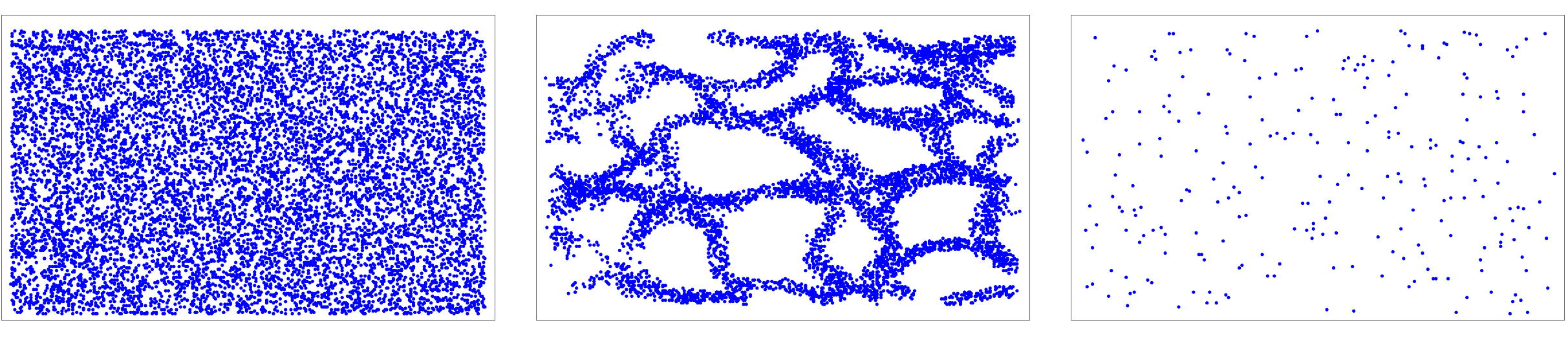}
\caption{A representation of hypothesized DM distribution given in three panels. {\it Left panel} shows the initial homogeneous distribution, possibly for an uncoupled DM-plasma. {\it Middle panel} represents the era, where DM-structures dominate the universe. {\it Right panel} is when the structures break apart and self-interaction effects cease to exist. (Figures are not derived from actual data or simulations. They are generated for illustrative purposes only.)} 
\label{evolution} 
\end{figure*}

Model examined in the previous section is valid for static fluids and therefore only applicable for a limited timeframe in cosmology. However it is necessary to construct a model for DM self-interactions (DMSI) as a function of redshift so that we can compare it to cosmological datasets. We assert that those interactions would start to affect cosmic expansion after a hypothetical DM-plasma decoupling and start to increase as cosmic structures evolve (such an increasing energy density would be {\it phantom} by definition). It is also plausible to think that they will start to lose their strength after {\it pinching} of DM walls and filaments when the universe is diluted enough.  (for a graphical representation see Figure \ref{evolution}). This is also in line with the big rip scenarios, in which all structures within a phantom dominated universe would be thorn apart in finite time. In a tension-driven acceleration model, aforementioned structures are also the cause of the rip.

To construct a cosmological model that takes DMSI into account, one should consider a parameter space of DM-particle masses and interaction cross-section, as well as DM density (a model-dependent parameter estimated from cosmological data). The size of the parameter space itself would depend on the complexity of the model; number of DM-particle species, type of interaction(s), etc. Moreover, cosmological features mentioned above, that would weaken and stop DMSI effects in finite future, would not be emergent from particle physics approach alone.

Nonetheless, we would like to examine the cosmological characteristics of such model. Therefore we take an alternate approach and come up with a phenomenological model for this DMSI-driven ``energy'' density, that would be nonexistent in the far-past, increase as DM structures grow and start to diminish after reaching a maximum, in the near-past or future. To do so we propose a simplistic expression, basically for the density $\mu$ in (\ref{rhos}), which is the multiplication of two power functions of redshift (common for every perfect fluid in cosmology), one for the increasing density era, $(10^3-z)^{\beta}$ and the other for the decreasing, $(z+1)^{\alpha}$, with necessary conditions, $0<\alpha<1$ and $1<\beta$:

\begin{equation}
\rho_{\rm DMSI}(z)=\rho_0(z+1)^{\alpha}(10^3-z)^{\beta}10^{-3\beta} 
\end{equation}

Our assumption in this model is that the type of energy content will have zero effect at around $z=1000$ when the universe was significantly homogeneous and at $z=-1$ when the expansion goes to infinity. The last factor is introduced as a ``$z=0$" correction (There are also different phenomenological models for dark energy density in literature with different motivations \cite{li2019simple}). The luminosity distance function for this model, assuming zero spatial curvature, would be given as,

\begin{equation}
d_L(z)=\frac{(z+1)}{H_0}\int \frac{10^{3\beta} c dz}{\sqrt{\Omega_{\rm M}+\Omega_{\rm DMSI}(z+1)^{\alpha}(10^3-z)^{\beta}}}
\end{equation}

where $\Omega_{\rm M}$ is the fraction of the matter density (baryonic or dark) and since assuming flatness, contribution from DMSI is $\Omega_{\rm DMSI}=1-\Omega_{\rm M}$. We neglected the radiation component because we are only interested in late-time cosmological data, i. e. Type Ia supernovae (SNe Ia) and quasars. 

We use the recent Pantheon compilation \cite{scolnic2018complete} to estimate the model parameters via standard $\chi ^2$ minimization. Best values are given in Table \ref{parameters}. The model with DMSI fits the data slightly better than $\Lambda$CDM.

\begin{table}
\caption{Parameters for DMSI driven accelerated expansion model and $\Lambda$CDM. $H_0$ values are calculated for SNe Ia absolute magnitude interval of $19.2<M<19.3$ from \cite{richardson2014absolute}}
\centering
\begin{tabular}{@{}*{7}{l}}
\hline
& $\alpha$ & $\beta$ & $\Omega_{\rm M,0}$ & $H_0$ & $\chi^2$/dof\\
\hline
DMSI & 0.03 & 600 & 0.35 & 72.4-76.1 & 1021.28/1044 \\
\hline
$\Lambda$CDM & - & - & 0.29 & 71.9-75.8 & 1024.35/1046 \\
\hline
\end{tabular}
\label{parameters}
\end{table}

The obtained value 600 for $\beta$ may seem unusual for a parameter that should also be estimated from fundamental physics; however, this is only a byproduct of standard redshift parametrization. For example, using $y$-redshift parametrization introduced in \cite{cattoen2007cosmography}, $\beta$ would be around 1 for the same ansatz. 

\begin{figure}
\centering
\includegraphics[width=\columnwidth]{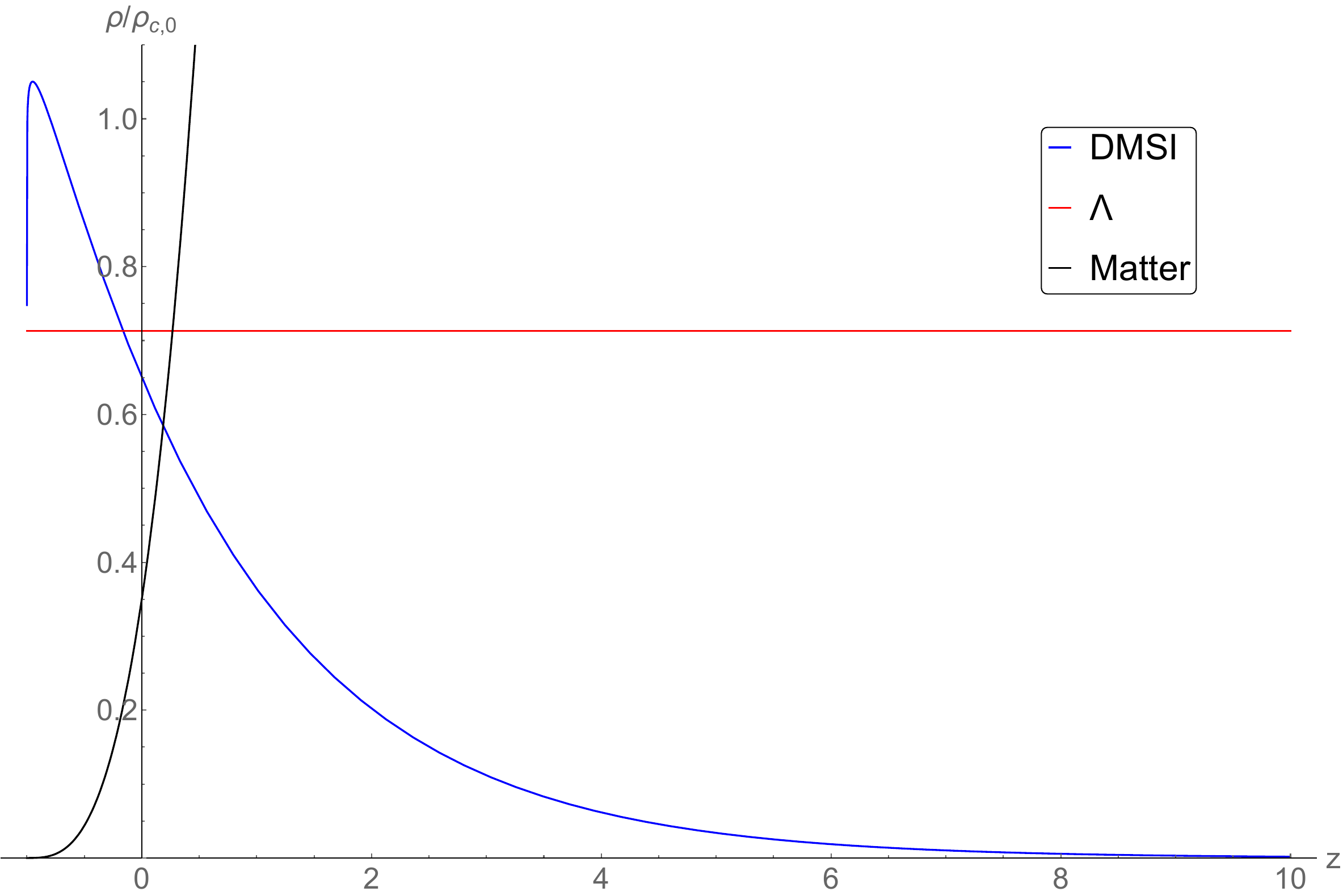}
\caption{Energy densities as a function of redshift for DMSI with $\alpha=0.03$ and $\beta=600$ (blue), $\Lambda$ (red), and matter (black).} 
\label{densities} 
\end{figure}

Matter-dark energy equality is shifted towards more recent times for the introduced model with respect to $\Lambda$CDM, as expected, and phantom behavior continues past $z=0$ (Figure \ref{densities}). Parameter $\alpha$ (``future" side of the curve) is lightly constrained by data. However, we see that DM-interactions dominate the cosmic expansion up to the point where the universe becomes 20 times its current size and then lose their effect. Therefore we conclude that data does not favor a turning point in the DMSI in the near past.

\begin{figure}
\centering
\includegraphics[width=\columnwidth]{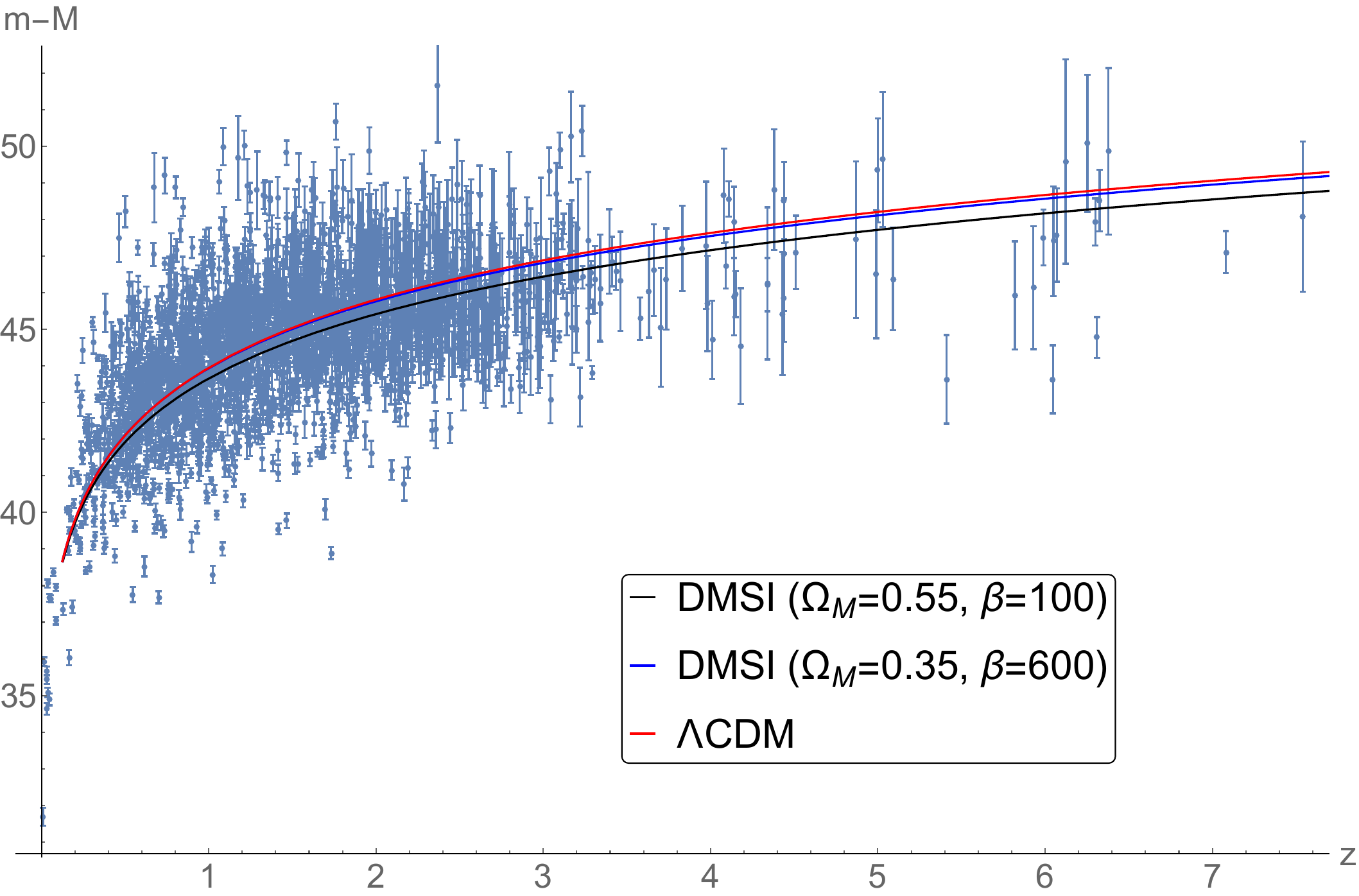}
\caption{Distance modulus - redshift diagrams for $\Lambda$CDM (red) and two different DMSI models (blue and black), alongside with quasar data. Parameter $\alpha$ was kept at 0.03, since its effect at redshifts beyond $z\sim1$ is negligible.} 
\label{quasar} 
\end{figure}

It is yet difficult to distinguish between two models statistically for the redshift range and precision of SNe Ia data. However, two energy densities differ more drastically for higher redshifts (Figure \ref{densities}). To see if there is a meaningful distinction between two models we include high redshift quasar data that are recently presented \cite{lusso2020quasars}. Comparing two different DMSI models with $\Lambda$CDM, we see that distance modulus curves for the new ansatz are significantly lower than that of $\Lambda$CDM for high redshifts (Figure \ref{quasar}). That feature, somewhat expected for a model with higher matter density and phantom behavior to compensate, is also what quasar data favor as seen in cosmographic analyses \cite{lusso2019tension}. This new observational inconsistency of 4$\sigma$ between $\Lambda$CDM and quasars, dubbed ``quasar Hubble diagram tension" \cite{perivolaropoulos2022challenges}, can be elevated by the proposed transient behavior, naturally.

With the interval for $\Omega_{\rm DMSI}$ inferred from data we can estimate for the actual value of $\sigma$ as it is directly related to $\rho_s$ by (\ref{foampress}), such that:

\begin{equation}
\sigma=\frac{\Omega_{\rm DMSI}\rho_c R}{3}
\end{equation}

Once again, assuming $H\sim70$ km/s/Mpc, and the average void radius $R\sim 100$ Mpc, we find the surface tension in DM-walls lies in the range:

\begin{equation}
\sigma \in [3.8\times10^{14}J/m^2\sim5.5\times10^{14}J/m^2] 
\end{equation}

When compared to the estimated surface tension from gravitational interaction (equation \ref{rhos}), $\sigma \simeq 1.2\times10^{11}J/m^2$, we see that DM-tension is more than $10^3$ times larger, in agreement with the estimation given at the end of the previous chapter.

This large value of surface tension may lead us to think about the stability of DM-structures proposed in this model; would they collapse under such forces or not? In order to discuss the stability of those structures, first we have to keep in mind that this surface tension, as proposed by the model, is not constant in time and increases from a near-zero value as the structures form from an effectively homogeneous background. Secondly, tension by definition, occurs only if forces in the outward direction are present and self-interactions in our model are opposing them. As assumed in our model, voids are empty and pressureless, so this effective outward pull on the structures is driven by the expansion itself. Therefore our scenario does not directly foresee a collapse of the voids, but rupture of the walls in the future are more likely. The increase in $\sigma$ in that manner can also be understood as self-interacting DM particles (compact/bound DM object or DM substructures, as well) being redistributed toward a configuration of higher potential energy as space expands.

\section{Conclusion}

Works that relate self-interactions of DM to the late acceleration of the universe exist in the literature \cite{atreya2018viscous}; however, with no emphasis on structure formation as a triggering mechanism. There are also works that suggest an acceleration driven by collapse and tension in the emerging structures \cite{yusofi2022surface}. It was also argued that surface tensions originating from Newtonian gravity can cause accelerated expansion \cite{ortiz2020surface}. We beg to differ the results in \cite{yusofi2022surface} and \cite{ortiz2020surface} that a Newtonian tension would naturally compensate for an observed acceleration as our estimation suggested. Our main difference with \cite{yusofi2022surface} is that, even though relying on similar principles, we were able to estimate deceleration parameter using global parameters like Hubble expansion rate or average void radius (equation \ref{deceleration}), whereas \cite{yusofi2022surface} takes mass density of the structures into account, a quantity that varies dramatically for different scales. Additionally we think that those effects should be negligible in the past and diminish in the far future as they do in our phenomenological model. 

It is too early to speculate on the type or strength of the DM self-interactions; we are still far from telling if they even exist or not (for a discussion of astrophysical effects of DM-interactions see \cite{foot2016solving}). But we can lay down a framework for our cosmic scenario assuming that DM is self-interacting:

First of all, the DM self-interaction should be strong enough --maybe on the electromagnetic scale-- to support high tensions that can cause negative deceleration. Secondly, formed DM structures should not be neutral, unlike structures bound under electromagnetism, or they should at least expose strong {\it van der Waals} type leaks, to reach intergalactic scales, as gravity does. Additionally, DM should stay in a plasma state up until late times, at least later than photon-baryon decoupling, in accordance with the beginning of the acceleration epoch. Lastly, a fast condensation reminiscent of a phase transition or a more complex {\it chemical} interaction picture that results in strong bonds between DM-particles and substructures may be needed to avoid early pinching of the cosmic DM-filaments/walls. 

The effects of dark matter on large scale structure formation are investigated through cosmological observations with indicators of those effects are being forecast using N-body simulations. This area of research, however, still depends on many uncertainties and is not yet predictive enough to favor one dark matter model over another. The most simplistic scenario, where initial noninteracting dark matter inhomogeneities lay the foundation for the matter structure formation is still the benchmark theory, however it is to be confirmed if dark matter inhomogeneities and matter structures are spatially correlated and exhibit similar power spectra \cite{angulo2013closely}. N-body simulations work very well for a universe consisting of dust particles under general relativity. But when the dark matter interactions are considered their predictive power becomes more questionable. Incorporating self-interacting dark matter scenarios to the N-body simulations is challenging, since many factors like type of interaction (dissipative, decaying or totally self-interacting), interaction cross-section, interaction strength and magnitude and starting redshift for the initial perturbations in the simulation play a significant role. Choice of different scenarios and parameters may lead to a variety of observational differences, but may result in similar power spectra as well \cite{angulo2022large}. The most important parameter, tension strength $\sigma$ in our model represents an average overall effect, that may arise from a combination of aforementioned scenarios and parameter choices. As a future outlook, it is possible to construct an effective model for the power spectrum of dark matter structures, using $\sigma$ and $\rho_s(z)$, in order to compare it to the results from N-body simulations and cosmological observations.

From a field theoretical perspective, one may also want to extend the model by including other types of interactions and fields and calculate their effects from the first principles, as is done in the mean field theories. To do so, one needs a covariant averaging scheme for cosmology as a whole. While spatial averaging of noninteracting matter inhomogeneities has been studied in literature -- leading to the concept of cosmological backreactions, which may contribute to the observed accelerated expansion of the universe \cite{schander2021backreaction} -- a fully covariant space-time averaging remains significantly more challenging. Despite the most profound attempts, the averaging problem in cosmology persists \cite{clarkson2011does}. In this work we adopted a top-down approach, suggesting a model for self-interactions in DM-structures described by a single overall tension that changing over time and considered consistent cosmic scenarios and constraints from this framework.

An increased number of constraints, in this case, does not necessarily mean that we are dealing with a more complex and fine-tuned model. One should keep in mind that models that include dark energy still include dark matter, maybe an already self-interacting one. We have argued that, if this interaction has certain properties, the apparent acceleration may be explained without the need for dark energy.

We depend on future observations to see if this mechanism is viable within a reasonable interaction picture. In the meantime, computer N-body simulations, running on different types of DMSI models, would be the way to get the most out of this scenario and to see if a strong enough mechanism can be found to drop dark energy from the cosmic picture; to be replaced with particle interactions, a more familiar and natural, less exotic concept.

\section{Acknowledgements}  Author would like to thank Dr. Ne\c se Aral for her help with the first figure.

\section{References}

\bibliographystyle{unsrt}
\bibliography{sidmbib}

\end{document}